\documentstyle[12pt]{article}

\def\be{\begin{equation}}
\def\ee{\end{equation}}
\def\bea{\begin{eqnarray}}
\def\eea{\end{eqnarray}}
\def\ba{\begin{array}}
\def\ea{\end{array}}
\def\ben{\begin{enumerate}}
\def\een{\end{enumerate}}

\def\nnu{\nonumber}
\def\ll{\label}
 
\begin{document}
\newcommand{\half}{{\textstyle\frac{1}{2}}}
\newcommand{\eqn}[1]{(\ref{#1})}
\newcommand{\npb}[3]{ {\bf Nucl. Phys. B}{#1} ({#2}) {#3}}
\newcommand{\pr}[3]{ {\bf Phys. Rep. }{#1} ({#2}) {#3}}
\newcommand{\prl}[3]{ {\bf Phys. Rev. Lett. }{#1} ({#2}) {#3}}
\newcommand{\plb}[3]{ {\bf Phys. Lett. B}{#1} ({#2}) {#3}}
\newcommand{\prd}[3]{ {\bf Phys. Rev. D}{#1} ({#2}) {#3}}
\newcommand{\hepth}[1]{ [{\bf hep-th}/{#1}]}
\newcommand{\grqc}[1]{ [{\bf gr-qc}/{#1}]}
 
\def\a{\alpha}
\def\b{\beta}
\def\g{\gamma}\def\G{\Gamma}
\def\d{\delta}\def\D{\Delta}
\def\ep{\epsilon}
\def\et{\eta}
\def\z{\zeta}
\def\t{\theta}\def\T{\Theta}
\def\l{\lambda}\def\L{\Lambda}
\def\m{\mu}
\def\f{\phi}\def\F{\Phi}
\def\n{\nu}
\def\p{\psi}\def\P{\Psi}
\def\r{\rho}
\def\s{\sigma}\def\S{\Sigma}
\def\ta{\tau}
\def\x{\chi}
\def\o{\omega}\def\O{\Omega}
\def\k{\kappa}
\def\pa {\partial}
\def\ov{\over}
\def\br{\nonumber\\}
\def\ud{\underline}
\begin{flushright}
DFPD/00/Th/12\\
SINP/TNP/00-06\\
hep-th/0003106\\
\end{flushright}
\bigskip\bigskip
\begin{center}
{\large\bf 
Action with manifest duality for maximally supersymmetric six-dimensional
supergravity}
\vskip .9 cm
{\sc Giancarlo De Pol$^a$\footnote{e-mail: depol@pd.infn.it}, 
Harvendra Singh$^b$\footnote{e-mail: hsingh@tnp.saha.ernet.in},
Mario Tonin$^a$\footnote{e-mail: tonin@pd.infn.it}}
\vskip1cm
$^a$ Dipartimento di Fisica `Galileo Galilei', INFN Sezione di Padova, \\
Universit\`{a} di Padova, Via F. Marzolo 8, 35131 Padova, Italia
 \vskip 0.5cm 
$^b$ Theory Division, Saha Institute of Nuclear Physics,\\
 1/AF Bidhannagar,
 Calcutta-700 064, India 

\end{center}
\bigskip
\centerline{\bf ABSTRACT}
\bigskip

\begin{quote}
We perform explicitly the toroidal compactification of eleven
dimensional supergravity to six dimensions  and present its action
in a manifestly
$SO(5,5)\over SO(5)\times SO(5)$ invariant form using
the recently proposed covariant formulation of theories involving chiral
fields.
\end{quote}

\newpage
\section{Introduction}
The compactification of $d=11$ and $d=10$ supergravities to lower dimensions 
is an old subject that recently has acquired new interest, due to the
important r\^{o}le that dualities have reached in modern string 
theories and in M-theory. Even the simplest compactifications of $d=11$ 
supergravity  \cite{cre-jul-sch} on flat tori have received new interest. 
They give rise to 
lower dimensional supergravities with maximal supersymmetry and 
maximal duality groups (U-duality). The presence of these  hidden 
symmetries was recognized from the beginning \cite{sal-sez} and  they are by 
now  
completely classified \cite{jul} and widely investigated \cite{cre-jul-lu}-
\cite{nas-vam-nie}.
These maximal supergravity models are useful  
at least as a first step to study more elaborate (and more interesting) 
models derived from them. An example are the massive and gauged 
supergravities in different dimensions \cite{hs,vol}. 
Another related example is given by compactifications on spheres 
\cite{cve-liu-lu, nas-vam-nie}, now so 
popular after the advent of the AdS/CFT duality.
A characteristic feature of U-duality is that, in general, the bosonic 
multiplets of the duality group in $d$ dimensions contain both, let say,
 $p$-forms and the duals of $(d-p-2)$-forms. This happens because U-duality 
contains transformations which correspond to electromagnetic duality and 
are conjectured to arise at a non-perturbative level in string models.
In particular, the duality group of the maximal $d=6$ supergravity coming 
from the compactification of $d=11$ supergravity on a five-torus is 
$SO(5,5)$, and one must consider the five  2-form potentials which arise 
from the reduction of the 3-form potential in $d=11$, together with their 
duals potentials, to fill the ten-dimensional vector multiplet of 
$SO(5,5)$.                              
The maximal $d=6$ supergravity is well known and its action has been 
obtained long time ago in a beautiful paper  by Tanii \cite{tan}. This action
 has been derived from its field content and not 
as an explicit dimensional reduction from $d=11$ supergravity.
Moreover, the $SO(5,5)$ duality in the sector of the 2-form potentials is 
manifest in the action of \cite{tan} only at the level of field equations. 
Recently a formulation of covariant actions with manifest 
duality has been available \cite{pst}.
In \cite{dal-lec-ton} - \cite{hoof} this approach has been applied to $d=6$ supergravity models.
In the present paper we obtain an action for the maximal $d=6$
supergravity by 
performing an explicit toroidal compactification from $d=11$ and, by using 
this  
approach, we present this action in a form that is 
covariant and manifestly invariant under the $SO(5,5)$ duality group. 
Our initial motivation to consider this case, was just our wish to use the 
approach of ref.\cite{pst} at work in a simple case, as a first step to study 
more interesting situations. However, in this program we met some technical 
subtleties that convinced us to write this paper. 
For instance, the one-form potentials in $d=6$ belong to a Weyl-Majorana 
spinor of $SO(5,5)$, and in order to write the action 
terms involving their field strengths in a form with manifest $SO(5,5)$ 
invariance, one needs a vielbein-like matrix which belongs to the coset
$SO(5,5)/ (SO(5)\times SO(5))$. However, the construction of this matrix
 in terms of the scalars that arise from the reduction is not 
straightforward.
\par
In the short section 2 we review briefly the $d=11$ supergravity 
\cite{cre-jul-sch}, writing its action and supersymmetry transformations, 
in order to set our notations. In section 3 we perform the reduction on 
$M_6 \times T_5$ of the bosonic sector of this action and express the reduced 
action in a covariant form which is manifestly invariant under $SO(5,5)$ 
duality. In particular we give the explicit form of vielbein-like matrix 
$V$ in terms of the scalars coming from the reduction. In section 4 we 
describe the reduction of the fermions in order to complete the action and
write the supersymmetry transformations of the $N=4$ , $d=6$ supergravity 
(however, the four-fermion terms have not been worked out explicitly).  

\section{11-dimensional Supergravity}

In this section we set our notations and briefly review some elements of
the eleven dimensional supergravity. We use mostly plus signature
for the spacetime metric. The latin alphabet is employed for vector
indices while greek letters are used to describe spinor indices, but
the last will be often suppressed.
In our convention, a $p$-form and its Hodge-dual in $d$-dimesional space of 
signature $t$ are defined by
\bea
&&V_{(p)}={1\ov p!} \ e^{a_1}\cdots e^{a_{p}} V_{a_p\cdots a_1}\\ 
&&^\ast V_{(p)}={1\ov (d-p)!p!} \ e^{a_1}\cdots e^{ a_{d-p}}
\epsilon_{ a_{d-p}... a_1}\,^{ b_1... b_p}V_{
b_p\cdots b_1} 
\eea
so that $^{\ast\ast} V_{(p)}=(-)^{t+d(d-1)/2}~V_{(p)}$.
Here, $ e^{ a}$ are one-form vielbeins, $ \g^{ a_1\cdots a_p }$
 denotes the 
antisymmetric product of $p$ Dirac matrices $\g^{a} $, normalized to $1$,
$$\g^{(p)}={e^{a_1}\cdots e^{a_p}\ov p!}~\g_{a_1\cdots a_p}$$ 
and $\G^{(p)}=C_{(d)}\g^{(p)}$ , where $C_{(d)}$ is the charge 
conjugation in $d$ 
dimension. Of course, $V_{(p)}V_{(q)}=(-1)^{pq}V_{(q)}V_{(p)}$  is the
wedge product of differential forms. Objects in eleven dimension will be
marked by a hat.

The standard supergravity \cite{cre-jul-sch} action in eleven dimensions 
is (for $\kappa=1$)

\bea
S_{(11)}=&&\int \bigg[ {\hat e \hat R\ov 4}-{1\ov
2}\hat G_{(4)}~^\ast\hat G_{(4)} 
-{1\ov3}\hat G_{(4)}\hat G_{(4)}\hat C_{(3)}
-{1\ov2}\hat\Psi\hat\Gamma^{(8)}\hat D\hat\Psi \br &&
-(\hat Q_{(7)}+^\ast\hat Q_{(4)}) \hat G_{(4)}
-{1\ov 2}(\hat Q_{(7)}+^\ast \hat Q_{(4)}) \hat Q_{(4)}
+{1\ov8}\hat\Psi\hat\Gamma^{(8)}\hat Q_{(1)}\hat\Psi
\bigg] 
\label{11}
\eea
where $\hat e=det(\hat e_{\hat m}^{~\hat a})$, $\hat R$ is the scalar
curvature,  
$\hat\P=\hat e^{\hat a}\hat\Psi_{\hat a}$ is a one-form Majorana
 spinor that represents the gravitino and $\hat C_{(3)}$ is a 3-form gauge
potential with curvature $\hat G_{(4)}=d \hat C_{(3)}$. Moreover
\bea
&&\hat D\hat\P= d\hat \P-{1\ov4}\omega_{\hat a \hat b}\hat\gamma^{\hat b \hat
a}\hat\P \\ &&
\hat Q_{(4)}={1\ov 4}\hat\P\hat\G^{(2)}\hat\P\\ &&
\hat Q_{(7)}= - {1\ov 4}\hat\P\hat\G^{(5)}\hat\P\\ &&
\hat Q_{(1) \hat\a}~^{\hat\b}=-{1\ov 8}\hat e^{\hat
c}(\hat\P_{\hat f}~\hat\G^{\hat c\hat b\hat a \hat f\hat g}~\hat\P_{\hat
g})(\hat\g_{\hat a\hat b})_{\hat\a}^{~\hat\b} .
\eea

and the supersymmetry transformations are
\bea
&&\d \hat e^{\hat a}= \hat{{ \ep}}~\G^{\hat a}\hat \P\\
&&\d\hat C_{(3)}= -{1\ov 2}\hat{\ep}~\hat\G^{(2)}\hat\P\\
&&\d\hat\P=\hat D\hat\ep -{1\ov 4}\hat Q_{(1)}\hat\ep\\ &&\hskip1.5cm
 -{1\ov {3!4!}}\left(\hat G_{\hat b_1...\hat b_4}-\hat Q_{\hat b_1...\hat
b_4}\right)\left(\hat e^{\hat a} \G_{(1)}^{\hat b_4...\hat b_1\hat a}+8 
\hat e^{\hat b_4}\G^{\hat b_3\hat b_2\hat b_1}\right)\hat{\ep}     .
\eea
where $\hat{\ep}$ are the local supersymmetry parameters (a real spinor in 
$d=11$).

\section{Reduction on ${\cal M}_6 \times T^5$}

Let us consider the reduction of the $d=11$ spacetime to a six dimensional 
manifold ${\cal{M}}_{6}$ and a $5$-torus $T_5$.
The spacetime coordinates $\hat{x}^{\hat{m}}$ split up into two sets.
For the ones of  ${\cal M}_6$ we shall use $x^m,~(m,n,...)$,
without hat, and for
those of the five-torus we shall use $y^u,~(u,v,...)$.
The corresponding
tangent space indices are $(a,b,...)$ and $(i,j,...)$ respectively.
The ansatz for the vielbeins is  
\be
\hat e^{\hat a}= \left( \o^{-1/6}e^a,~~\o^{1/3} \eta^u e_u^{~i}\right)
\ll{21}
\ee
where
$\o^{-1}=det (e_u^{~i})$ and the metric on the internal space is 
$g_{uv}=e_u^{~i}\d_{ij}e_v^{~j}$. The 
one-form $\eta^{u}$ is $\eta^u=dy^u+A^u$ where
$A^u$ are the Kaluza-Klein gauge fields.

It is now straightforward to work out the reduction of the Einstein term
in the action \eqn{11}. One finds that

\be
\int{1\ov4}\hat e \hat R = \int{1\ov4 }\left(
e R +{1\ov 4} ~dg^{uv}~^\ast dg_{uv} + {1\ov 2}\o
F^u~^*F^v g_{uv} \right)
\ll{22} \ee
where $F^u=dA^u$. 
One must note that
the Einstein term on the  right hand side of eq.\eqn{22} is independent
 of $\o$ which is due to 
judicious choice of scaling factors in the ansatz \eqn{21}.  
\vskip.5cm 
\noindent{\underline{\it Reduction of $\hat G_{(4)}$}}

The 3-form potential reduces as follows :
$$
\hat C_{(3)}=  C'_{(3)}+dy^u B'_{(2)u}+{1\ov 2}dy^u dy^v 
A'_{(1) vu}+{1\ov6}dy^udy^vdy^w \f_{(0)wvu},$$
The forms on                  
the r.h.s. of the above equation do not depend upon the coordinates $y^u$
of the torus. The reduction of corresponding 
 4-form field strength can be written as
\be
\hat G_{(4)}=
G_{(4)}-\eta^u H_{(3)u}+{1\ov2}\eta^u\eta^v (F_{(2)vu}+\f_{(0)vuw}F_{(2)}^w)
-{1\ov6}\eta^u\eta^v\eta^w S_{(1)wvu}
\ll{23}\ee
where (dropping the suffixes $(p)$ that mark the form degree)   
\bea
&&S_{uvw}= d\f_{uvw},~~~F_{uv}= dA_{uv},\br
&&H_u=d B_u-{1\ov 2}(A_{uv}F^v + F_{uv}A^{v}),\br
&&G= d C +B_uF^u-{1\ov 2}A_{uv}A^vF^u,
\ll{24}\eea
and
\bea
&& A_{uv}= A'_{uv} -\f_{uvw}A^w\br     
&& B_u=B'_{u} -{1\ov 2}A_{uv}A^v-{1\ov2}\f_{uvw}A^wA^v,\br
&& C= C'-B_{u}A^{u}-{1\ov6}\f_{uvw}A^wA^vA^u.
\ll{26}\eea
Then the kinetic term for the
3-form potential $\hat C$ will reduce to
\bea
-{1\ov2}\int \hat G~^\ast\hat
G&=&-{1\ov2}\int\bigg[\o^{}G~^*G - H_u~^*H^u
+ {\o^{-1}\ov2}(F_{uv}+\f_{uvw}F^w) ~^*(F^{vu}+\f^{vuw}F_w)
\br &&-{\o^{-2}\ov6}S_{uvw} ~^*S^{wvu}\bigg]
\ll{27}\eea 
and the Chern-Simon term becomes
\be
\ll{28}
-{1\ov3}\int\hat C\hat G\hat G=
\int\left(G~{F}_{vu}\F^{uv}-
G~{ F}^z{\tilde\f}_{z}+H_vH_u\F^{uv}\right) +
\int{\ep^{uvwxy}\ov4}{ F}_{yx}{ F}_{wv} (B_{u}+{1\ov 2}A_{uz}A^{z}) 
\ee 
where we have defined
\be
\ll{29}
\F^{uv}={1\ov3!}\ep^{uvwxy}\f_{yxw},~~~{\tilde
\f}_v=-{1\ov2}\f_{vxy}\F^{yx}.
\ee
Also let us write the interaction term between fermions and $\hat G$ as
\be
\ll{31}
-\int(\hat Q_{(7)}+^*\hat Q_{(4)})\hat G=\int\left(G~^*Q_{(g)}+H_u
Q_{(h)}^u+(F_{uv}+\f_{uvw}F^w)Q_{(f)}^{vu}+S_{uvw} Q_{(s)}^{wvu}\right)
\ee
where $Q_{(g)},~ Q_{(h)}^{u}$ etc. are forms bilinear in the fermions
whose specific form will be determined later.

Since the reduced action is at most quadratic in $G$, one can dualize $C$
in standard way: one lets G to be generic, adds the term 
$(G-(B_u-{1\ov 2}A_{uv}A^{v})F^u)dA^{\otimes}$ (where $A^{\otimes}$ is a 
Lagrange multiplier 1-form) and
integrates over $G$. Then the part of the
action involving $G$ can be replaced by the dual action 

\bea\ll{34}
S_{dual}&=&\int\bigg[ {1\ov 2}\o^{-1} ({F^{\otimes}}+{F}_{vu}\F^{uv}-
{ F}^z{\tilde\f}_{z}+~^*Q_{(g)})~^*({F^{\otimes}}+{
F}_{vu}\F^{uv}- {F}^z{\tilde\f}_{z}+~^*Q_{(g)})\br && -({
B}_u-{1\ov 2}A_{uv}A^{v}) F^uF^{\otimes}\bigg] 
\eea 
where $ F^{\otimes}=dA^{\otimes}$ is the field strength of the one-form
${A^{\otimes}}$. One can
 note that the equation of motion for $A^{\otimes}$ exactly reproduces
the Bianchi identity $dG=H_u~F^u$ for its dual \mbox{3-form $C$.}
\par
The gauge transformations of the potentials $A^u$,$~A_{uv}$,$~A^{\otimes}$
are
\bea 
&& \d A^u=d \l ^u \nnu\\
&& \d A_{uv}=d \l_{uv}\nnu\\
&& \d A^{\otimes}=d \l^{\otimes}
\ll{gaugetr}
\eea
and since $H_u$ must be gauge invariant it follows from eq.\eqn{24} that 
also $B_u$ transforms under \eqn{gaugetr} as 
\be
\d B_u = {1\ov{2}}(\l_{uv} F^v + F_{uv} \l^v) .
\ll{gaugeb}
\ee 
Moreover $B_u$ transforms under its own gauge transformation: 
$\d B_u = d\L _u$ where $\L_u$ is a one-form.
\par
Let us count the total number of 1-form potentials after reduction.
All 1-form gauge fields $A^u$, $A_{uv}$, $A^{\otimes}$ 
add up to sixteen, which can be arranged into
a Majorana-Weyl representation of the duality group $SO(5,5)$. Before 
dealing with it, we must explain our notations for $SO(5,5)$ in next
subsection.
\vskip.5cm
\noindent{\underline{\it Action for 1-form gauge fields}}

The spinorial representation for $SO(5,5)$ is
32-dimensional. We denote
vector indices of $SO(5,5)$  by capital letters $R,S,...$ and
a vector $v^R$ is composed of two five-dimensional parts
$v^R=(v^{(1)r},v_r^{(2)});~r=1,...,5$. The metric is off-diagonal
$\eta_{RS} v^Rv^S= (v^{(1)r}v^{(2)}_r+v^{(2)}_s v^{(1)s})$. The
$\g$-matrices can be conveniently constructed
as a tensor product of two $SO(5)$ $\g$-matrices and Pauli matrices:  
$$\g^r\times I_4\times \sigma_1, ~~I_4\times\g^r\times(i \sigma_2)$$
and in the chosen metric they are
\be\ll{43cc}\g^R=(\g^{(1)r},~\g^{(2)}_r)
\ee
where
\bea\ll{43ce}     
&&\gamma^{(1)r}={1\ov\sqrt{2}}\left(\g_{r}\times I_4\times\s_1 +
I_4\times\g_{r}\times i\s_2\right)
\br &&\gamma^{(2)}_r= {1\ov\sqrt{2}}\left(\g_{r}\times
I_4\times\s_1-I_4\times\g_{r}\times i\s_2\right).
\eea

In $SO(5,5)$ there exist Majorana-Weyl spinors with 16 components
$\psi_{\ud\m}\equiv
\psi_{\m}^{~\dot\m}$;\mbox{$~{\ud\m}=1,...16$}, \linebreak
\mbox{$~\m,\dot\m=1,...4$}
that satisfy the charge conjugation condition \footnote{More precisely 
$\psi=\pmatrix{\psi_{\ud{\m}}\cr 0}$ .}
\be\ll{43b}
{\psi}^c_{\ud\m}=(C^{-1})_{\ud\m}~^{\ud\n}{\bar\psi}_{\ud\n}=\psi_{\ud\m}.
\ee
Now let us come to the reduced action. Keeping from eqs.\eqn{22},
\eqn{27}, \eqn{34} 
the terms quadratic in the $F$'s one gets the kinetic action:
\bea\ll{43c}     
S_A&=&\int\bigg[ {1\ov 2}\o^{-1} (F^{\otimes}+{F}_{vu}\F^{uv}-
{ F}^z{\tilde\f}_{z})~^*(F^{\otimes}+{
F}_{vu}\F^{uv}- {F}^z{\tilde\f}_{z}) 
\br&& - {{\o}^{-1}\ov4}(F_{uv}+\f_{uvw}F^w)^*(F^{vu}+\f^{vuw}F_w)
+{\o\ov8}F^u~^*F_u \bigg]
\eea
As already noted, the sixteen 1-form potentials ($A^{\otimes},~A^u,~
A_{uv}$) can be arranged into a Weyl-Majorana $SO(5,5)$ spinor of,
say, positive chirality; this spinor is

\be\ll{44}
{\cal A}_{\ud\m}=A^{\otimes}\d_{\m}~^{\dot\m} +{1\ov2}A^u(\g_{\ud
u})_{\m}~^{\dot\m}
+{1\ov2}{A}_{uv}({\g^{\ud{u}\ud{v}}})_{\m}~^{\dot\m}
\ee
where $\g_{\ud{ u}}$ with underlined vector indices are constant $SO(5)$
matrices, so that
\be \ll{fda}
{\cal F}_{\ud\m}=d{\cal A}_{\ud\m}
\ee
is also a Majorana-Weyl(W-M) spinor. To write an action invariant under
$SO(5,5)$ one needs a veilbein-like matrix 
$V_{\ud\a}~^{\ud\m}$ that acts as a bridge between $SO(5,5)$ and its maximal
subgroup $SO(5)\times SO(5)$. The index $\ud{\a}$ is $\ud{\a}\equiv
(\a \dot\a)$.
$V_{\ud\a}~^{\ud\m}$ transforms as a W-M spinor of
negative chirality under $SO(5,5)$ from the right and as spinors of the  two
$SO(5)$ from the left. The spinor indices of the two $SO(5),~\a$ and $\dot\a$,
can be raised and lowered with the metric given by the $SO(5)$ charge
conjugation. Then the action \eqn{43c} should be rewritten as
\be\ll{act2}
S_{A}= -{1\ov 2(16)}\int {\cal F}_{\ud\a} ~^*{\cal
F}^{\ud\a}\ee  
$$ {\cal F}_{\ud\a}= V_{\ud\a}~^{\ud\m}~
{\cal F}_{\ud\m}$$
which is manifestly  invariant under global $ SO(5,5) $ transformations as 
well as local
$ SO(5)\times SO(5 )$ rotations.
A suitable gauge choice of these local transformations should allow us to 
express
$V_{\ud\a}~^{\ud\m}~$ in terms of 25 scalar fields $g^{uv}$ and $\F^{uv}$,
and to make eq.\eqn{43c} to exactly reproduce \eqn{act2}. A convenient gauge
fixed choice for $V_{\ud\a}~^{\ud\m}$ is
\bea       
&& \ll{VXY} V=\exp{X} \cdot \exp{Y}\\
&&  \ll{X=} X= {1\ov 4} f^{\ud{u} \ud{v}} [
\g_{\ud{u} \ud{v}} \otimes I_4 \otimes I_2 -
I_4\otimes \g_{\ud{u} \ud{v}} \otimes I_2 - 
2 \g_{\ud{u}}\otimes\g_{\ud{v}} \otimes \sigma_3]\\
&&  \ll{Y=} Y= {1\ov 4} f^{\ud{u} \ud{v}} [
\g_{\ud{u} \ud{v}} \otimes I_4 \otimes I_2 +
I_4\otimes \g_{\ud{u} \ud{v}} \otimes I_2 ]+
{h^{\ud{u}\ud{v}}\ov{2}}(\g_{\ud{u}}\otimes\g_{\ud{v}} \otimes \sigma_3)
\eea
where $X,~ Y$ belong to $SO(5,5)$ Lie algebra and
$h^{\ud{uv}},~f^{\ud{uv}}$ are related to the scalars $g^{uv},~\F^{uv}$.Given 
the ``vielbeins''
$$ e^{~\ud{u}}_{u}=\{ exp(h+f)\}^{~\ud{u}}_{u}$$  and its inverse
$$ e_{~\ud{u}}^u=\{ exp-(h+f)\}_{~\ud{u}}^u,$$ one has
\be\ll{etag}
 e_{~\ud{u}}^u \eta^{\ud{uv}}e_{~\ud{v}}^v=g^{uv} 
\ee
and  
\be\ll{efie}
e^{u}_{~\ud {u}} f^{{\ud u}{\ud v}} e_{~\ud{v}}^v=\F^{uv}.
\ee
Here, $\g_u=e_u^{~\ud u} \g_{\ud{u}}$ and so on.
That the gauge fixed matrix $V_{\ud\a}~^{\ud\m}$ defined in \eqn{VXY} allows
eq.\eqn{43c} to coincide with eq.\eqn{act2} can be easily verified by
noting that $exp[Y]$ applied to ${\cal F}$ {\it dresses} the
constant $\g$ matrices in ${\cal F}$ as
\be\ll{45a}
\big((exp Y)~{\cal F}\big)_{\ud{\s}}= {
F^{\otimes}}\o^{-{1\ov2}}\d_\s^{~\dot\s}
+{1\ov 2}F^u \o^{1\ov2}(\g_{u})_{\s}^{~\dot\s}
+{1\ov2}{F}_{uv}\o^{-{1\ov2}}({\g^{{u}{v}}})_{\s}^{~\dot\s}\equiv {\cal
F}^{(d)}_{\ud\s}
\ee
and
\bea\ll{45b}
\big((exp X)~{\cal F}^{(d)}\big)_{\ud{\a}}&&=  
\o^{-{1\ov2}}\left(
F^{\otimes} +F_{uv}\F^{vu}-F^u\tilde\f_u\right)\d_\a^{~\dot\a}
+{1\ov 2} \o^{1\ov2}F^u(\g_{u})_{\a}^{~\dot\a}\br &&
+{1\ov2}\o^{-{1\ov2}}\left(F_{uv}+F^w\f_{uvw}\right)
({\g^{{u}{v}}})_{\a}^{~\dot\a}~.
\eea
The matrix $V$ given above belongs to the coset
$SO(5,5)/ SO(5)\times SO(5)$ and transforms as
\be\ll{45c}
 V=h(g)Vg
\ee
where  $g$ is a spinorial transformation matrix of
$SO(5,5)$ and $h(g)$ is the compensating $SO(5)\times SO(5)$
transformation. The transformation \eqn{45c} implies the nonlinear
transformation of the scalars under $SO(5,5)$.
One can realize a 
partial gauge-fixing by allowing the coefficients $f^{\ud u \ud v}$ in
eqs. \eqn{X=} 
and \eqn{Y=} to be different and maintaining eq.\eqn{efie} only for the 
coefficients 
$f^{\ud u \ud v}$ in $Y$ (that is in eq.\eqn{Y=}). In that case the
action \eqn {act2} remains invariant under local transformations of the 
diagonal $SO(5)$
which corresponds to the Lorentz group for the compactified space.  

\vskip.5cm
\noindent{\underline{\it Action for B-fields}}\nopagebreak

Let us collect from eqs. \eqn{27}, \eqn{28}, \eqn{34} the terms 
involving the 2-form potentials together with those quartic in the gauge 
potentials. The action for the 2-form potentials is 
\bea\ll{35}
S_B&=&
\int\bigg[{1\ov2}H_u~^*H^u+H_u~ H_v\F^{vu}+H_u Q_{(h)}^u 
+{\ep^{uvwxy}\ov4}{ F}_{yx}{
F}_{wv}(B_u +{1\ov2}A_{uv}A^v)
\br &&-F^u F^{\otimes} (B_u - {1\ov{2}} A_{uv} A^v)\bigg]~.
\eea
The terms quartic in the gauge potentials are present in order to make 
eq.\eqn{35} invariant under the gauge transformations \eqn{gaugetr}  and \eqn{gaugeb}.
Let us rewrite $Q_{(h)}^u$ as \footnote{For the moment this splitting is 
quite arbitrary, but see below.}
\be\ll{QQQ}
Q_{(h)}^u=-Q^{(1)u}+ ^{*}Q^{(2)u} - 2 \F^{uv} Q^{(2)}_v
\ee 
and rename $B_u$ to $B_{u}^{(2)}$ to conform with our notation for vectors of 
$SO(5,5)$. Moreover, let us define its field strength as
\be\ll{hhq}
H^{(2)}_u=H_u + Q^{(2)}_u~ = dB_u^{(2)} + J_u^{(2)} + Q_u^{(2)}.
\ee
where
\be \ll{jfaf}
J^{(2)}_{u}={1\ov2}(F^w A_{wu}+A^w F_{wv})~.
\ee
Then eq.\eqn{35} yields the equation of motion 
\be\ll{36}
d\left[~^*H^{(2)u}-2\F^{uv}H^{(2)}_v-{1\ov4}\ep^{uvwxy}{ F}_{yx}{
A}_{wv}+
{1\ov {2}} (F^u {A^{\otimes}}+ A^u F^{\otimes})-Q^{(1)u}\right]=0 
\ee
while the Bianchi identity for $B^{(2)}_u$ is 
\be\ll{37}
d(H^{(2)}_u -J^{(2)}_{u}-Q^{(2)}_u)=0 .
\ee
If we  define
\be \ll{38}
H^{(1)u}=~^*H^{(2)u}-2\F^{uv}H^{(2)}_{v}
\ee
and
\be\ll{38a}
J^{(1)u}=
{1\ov4}\ep^{uvwxy}{A}_{yx}{ F}_{wv}-
{1\ov2}(A^u {F^{\otimes}}+ F^u {A^{\otimes}})
\ee
eq. \eqn{36} becomes
\be \ll{dhjq}
d(H^{(1)u} -J^{(1)u} -Q^{(1)u})=0 .
\ee
The closure of this 3-form allows us to define locally a 
2-form $B^{(1)u}$, dual to $B^{(2)u}$, such that
\be\ll{39}
H^{(1)u}= dB^{(1)u}+ J^{(1)u} +Q^{(1)u}.
\ee
In the notations introduced in eqs.
\eqn{43cc}, \eqn{43ce}, \eqn{44}, \eqn{fda},
equations \eqn{jfaf}, \eqn{38a} become 
\be
J^R = -{1\ov{16}} {\cal{A}}_{\ud {\m}} (\G^R )^{\ud{\m\n}} 
{\cal{F}}_{\ud{\n}}
\ll{jr16}
\ee
where $\G^R = C \g^R$, $J^R =\pmatrix{J^{(1)u}\cr J^{(2)}_u}$ and $C$ is the 
$SO(5,5)$ charge conjugation.
Eq. \eqn{jr16} shows that $J^R$ is an $SO(5,5)$ vector. 
We shall also write
\be
B^R=\pmatrix{B^{(1)u}\cr B^{(2)}_u} \nonumber,~~
H^R=\pmatrix{H^{(1)u}\cr H^{(2)}_u} \nonumber,~~
Q^R=\pmatrix{Q^{(1)u}\cr Q^{(2)}_u} \nonumber
\ee
so that 
\be
H^R = dB^R + J^R + Q^R
\ll{hdbjq}
\ee

We postulate that $B^R$ transforms
as a vector of $SO(5,5)$, and therefore 
$H^R$ also transforms as an $SO(5,5)$ vector, since, as we
shall show below, $Q^R$ is an $SO(5,5)$ vector, if the splitting in 
eq.\eqn{QQQ} is chosen appropriately.

The field equation and Bianchi identity \eqn{36}, \eqn{37} are
equivalently
described by the definitions \eqn{hhq}, \eqn{39}, \eqn{hdbjq} together
with the duality condition \eqn{38}.
Let us rewrite eq. \eqn{38} and its Hodge-dual as 

\bea\ll{40}
&& H^{(1)u}+2\F^{uv}H^{(2)}_v = ~^*H^{(2)u},\\
&& g^{uv} H^{(2)}_v = ~^*(H^{(1)u}+2\F^{uv}H^{(2)}_v).
\eea

Now let us introduce the matrix
\be\ll{matrixL}
L^I_{~R}=\pmatrix{L^i_u& L^{iu}\cr
\tilde{L}^{\tilde{i}}_u&\tilde{L}^{\tilde{i}u}}={1\ov\sqrt{2}}\pmatrix{
e^i_u & e^i_v(g^{vu}+2\F^{vu})\cr~ e^{\tilde{i}}_u & -e^{\tilde{i}}_v(g^{vu}-
2\F^{vu})}
\ee 
$ I=(i,\tilde{i}),~i,\tilde{i}=1,...5$, which can be written in more
compact form as
$$
L_{~R}^I={1\ov \sqrt{2}}\left(\begin{array}{cc}E &E^{-1T}+2E\F\\ E &
-E^{-1T}+2E\F \ea\right)~.
$$
The transposed matrix is 
$$ L_{R}^{~I}={1\ov{\sqrt{2}}}\left(\begin{array}{cc}E^{T}&E^{T}\\
{E^{-1}-2\F E^{T}}&{-E^{-1}-2\F E^{T}}\end{array}\right)~,$$
indices $I,J$ are raised (and lowered) with the metric 
$\tau^{IJ}=\left(\begin{array}{cc}I_5 & 0\\0 & -I_5\end{array}\right)$ 
(and its inverse).
$I_p$ stands for $p$-dimensional identity matrix, $E$ is the matrix with 
elements $e^i_{~r}$.
Some useful formulas are:
 
$$(L\eta L^T)^{IJ}\equiv\tau ^{IJ}$$ and 
$$(L^T\tau L)_{RS}=\eta _{RS}~,$$ where $$\eta _{RS}=
\left(\begin{array}{cc}0 & I_5\\I_5 & 0 \end{array}\right)_{RS}$$ is the 
$SO(5,5)$ metric. If , instead of $\tau$ , we use the metric $I_{10}$
\be \ll{43}  
M\equiv (L^T I_{10}L) =\pmatrix{G & 2G\F \cr -2\F G & G^{-1}-4\F~ G\F}~.
\ee
$G$ is the matrix
with elements $g_{uv}$ and $\F$ is the matrix with elements $\F^{uv}$. 

One can easily verify that  
\be\ll{45d}
L^I_R={1\ov32} Tr\left( \g^I V \g_R V^{-1} \right)
\ee
where, in terms of $SO(5)$~Dirac matrices, and in obvious notation,
$\g^I=( \g^i \times I_4 \times 
\sigma_1, 
 I_4 \times \g^{\tilde{i}} \times i \sigma_2$)~, 
while $\g^R$ is defined in \eqn{43cc}.
Eq.\eqn{45d} implies 
that $L^i_{~R}~(\tilde{L}^{\tilde{i}}_{~R})$ transforms as a vector of
$SO(5,5)$
from the right and as a vector of the first (the second) $SO(5)$ from the
left. To be more precise eq.\eqn{45d} leads to eq.\eqn{matrixL} if one uses
in it the partially gauge fixed $V_{\ud{\a}}^{~\ud{\m}}$ that preserves
local invariance
under the diagonal $SO(5)$  (the internal Lorentz group).
To get full local $SO(5) \times SO(5)$ invariance one must 
introduce in eq.\eqn{45d} the full ungauged $V_{\ud{\a}}^{~\ud{\m}}$ . In
that 
case eq. \eqn{45d} still reproduces eq. \eqn{matrixL} but now the vielbeins 
$e_u^{~i}$ and $e_u^{~\tilde{i}}$, even if constrained to give the same
metric $g_{uv}$, are different.
\par
If we define
\be \ll{42}
H^I=L_{~R}^I~H^R
\ee
that is $H^i=L^i_{~R} H^R,~ \tilde{H}^{\tilde{i}}=\tilde{L}^{\tilde{i}}_{~R}
H^R$, the sum and difference of eqs.\eqn{40} yield
\be\ll{433}
H^i=^*H^i,~~~\tilde{H}^{\tilde{i}}=-~^*\tilde{H}^{\tilde{i}}~.
\ee

Now it is straightforward to write down the action for the $B$ tensors using
the formalism in \cite{pst}, \cite{dal-lec-ton}. First we need to define
\bea
&&H_{-}^i=H^i-~^* H^i~~{\rm(self-dual)},\br&&
\tilde H_{+}^i=\tilde H^{\tilde{i}}+~^*\tilde H^{\tilde{i}}
~~{\rm(antiself-dual)}
\eea
and then we introduce an auxiliary 1-form $u$ 
\be\ll{43a}
u=da,~~~v={1\ov\sqrt{-u^2}}u
\ee
such that $v_mv^m=-1$, where $a(x)$ is an auxiliary scalar field. Using
the vector $v^m$ we can define the 2-forms \footnote{$i_v V_{(p)}$denotes the 
contraction of the $p$-form $V_{(p)}$ with the vector $v=v^m \partial_m$ .}
\be
{h}_{-}^i=i_{v}H_{-}^i,~~~\tilde{h}_{+}^{\tilde{i}}=i_v\tilde
H_{+}^{\tilde{i}}
\ee
One can also write
\bea
{h}^R
&=&i_v\left(L_{~~i}^R~ H_{-}^i+\tilde L_{~~\tilde{i}}^R~
\tilde H_{+}^{\tilde{i}}\right)
\br &=& L_{~~i}^R~{h}_{-}^i+\tilde L_{~~\tilde{i}}^R~
{\tilde{h}}_{+}^{\tilde{i}}~.
\eea
In a similar way we shall also write
\be\ll{jr}
J^R=L^R_i J^i + \tilde{L}^R_{\tilde{i}} J^{\tilde{i}}
\ee
and
\be\ll{qr}
Q^R=L^R_i Q^i + \tilde{L}^R_{\tilde{i}} Q^{\tilde{i}} .
\ee

Then \cite{pst},\cite{dal-lec-ton} the action for chiral fields
is
\bea\ll{pst}
S_B&&=\int \eta_{RS}\left(v ~h^R~H^S+{1\ov2}dB^R
(J^S + Q^S) +{1\ov2} J^R Q^S \right)
\br &&=\int \left(v ~h_{-}^i H_i +{1\ov2}H^i(J_i + Q_i) +{1\ov2}J^iQ_i \right)
-\left(v~\tilde h_{+}^{\tilde{i}} \tilde H_{\tilde{i}} +{1\ov2}\tilde
H^{\tilde{i}}
(\tilde J_{\tilde{i}} + \tilde Q_{\tilde{i}})
 +{1\ov2}\tilde{J}^{\tilde{i}}\tilde{Q}_{\tilde{i}} \right).
\eea
 One can see that the action
$S_B$ is invariant under the following transformations
\bea\ll{pst1}
&& i) ~\d B^R=d\L^R,~~~\d a=0\\
&&ii) ~\d B^R=-{{2h^R \xi}\ov\sqrt{-u^2}},~~~\d a=\xi \\
&&iii)~\d B^R=\zeta^R da,~~~\d a=0
\eea
where the two-form $\L^R$, the one-form  $\zeta ^R$ and $\xi$ are 
local parameters. The
transformation
$i)$ is the usual $B$-gauge transformation, invariance under $ii)$ implies
the auxiliary status of the scalar field $a(x)$ while the symmetry $iii)$
allows to
eliminate half of the propagating degrees of freedom carried by $B$.  
Indeed the field equations coming from the action \eqn{pst} are
\bea\ll{pst2}
&&d(v~h_{-}^i)=0\br
&&d(v \tilde{h}_{+}^{\tilde{i}})=0
\eea
and a suitable gauge fixing of the symmetry
$iii)$
together with the equations \eqn{pst2} leads to the  chirality conditions in
 \eqn{433}
$$H_{-}^i=0=\tilde{H}_{+}^{\tilde{i}}~.$$
Eq.\eqn{pst} is also invariant under the $A$-gauge transformations 
eqs.\eqn{gaugetr} which can be written as 
\be\ll{gaugeapst}
\d A_{\ud{\m}} = d \l_{\ud{\m}} .
\ee
It follows from \eqn{gaugeapst} and the transformation of $H^R$ that 
\bea\ll{deltajb}
\d J^R = -{1\ov{16}}d(\l_{\ud{\m}}(\G^R)^{\ud{\m\n}}
\cal{F}_{\ud{\n}})\nnu \\
\d B^R = {1\ov{16}}\l_{\ud{\m}}(\G^R)^{\ud{\m\n}} \cal{F}_{\ud{\n}}.
\eea
The invariance of $S_B$ under \eqn{gaugeapst} and \eqn{deltajb} is manifest 
except for the term ${1\ov{2}}\int{\eta_{RS} dB^R J^S}$. For this term it
follows from \eqn{deltajb} and the cyclic identity of $\G$ matrices in 
$10$ dimensions. Finally $S_B$ is manifestly invariant under $SO(5,5)$ and,
 trivially, under $SO(5)\times SO(5)$.

\vskip.5cm
\noindent{\underline{\it Action for scalar fields}

If we collect the bosonic terms involving scalars not yet considered, we get
\be
S_{scalar}={1\ov4}\int\left({1\ov4}dg^{uv}~^*dg_{vu}+ 
g_{uy}g_{vx}d\F^{uv}~^*d\F^{xy}\right) 
\ll{act3}\ee
Now it is straightforward to check that it can be written in a manifestly 
 $SO(5,5)$ invariant form
\be
S_{scalar}= {1\ov2~(16)}\int dM_{RS} ~^*dM^{RS} 
\ll{act3.1}
\ee
where $M$ is given in \eqn{43}. This action can also be presented in 
different equivalent forms.
For instance, let us define , in terms of the matrix $L$ \eqn{matrixL},
\be
P_{i\tilde{i}} = L^{i}_{~R} d\tilde{L}_S^{\tilde{i}}\eta^{RS} - 
dL^{i}_R \tilde{L}^{\tilde{i}}_{~S}\eta^{RS}
\ee
then it is straightforward to see that $S_{scalar}$ can also be rewritten as
\be\ll{haract}
S_{scalar}= -{1\ov 16}\int
P_{i\tilde{i}} ~^*P^{i\tilde{i}}.  
\ee
 
\section{Reduction of Fermions}
In the reduction of the eleven dimensional manifold to $M_6\times T^5$,
the $\g$-matrices in $d=11$ reduce as follows
$$ \g^{\hat a}=(\g^a,~\tilde{\g}^i{\bar \g})$$
where $\g^a~ (a=0,1,...5)$ are Dirac matrices in $d=6$ and
$\tilde{\g}^i~(i=1,...5)$ are Dirac matrices in $d=5$;
${\bar\g}=\g^0\g^1...\g^5$ is the chiral matrix in $d=6$. A Majorana 
spinor
in eleven dimensions is a spinor (32 complex components) that
satisfies the Majorana condition
$$\chi^c=C_{(11)}\bar\chi=\chi.$$
Its reduction to six dimensions gives us four symplectic W-M spinors with
positive chirality $\chi^\a_{+}$ and four  W-M spinors with
negative chirality $\chi_{-}^\a$, the symplectic condition being
$$\chi_{\pm\a}^c= (C_{(6)}\bar{\chi}_{\pm})_\a
=(C_{(5)})_{\a\b}\chi_{\pm}^\b.$$
The
fermions in $d=11$ are described by a 1-form Majorana spinor $\hat\Psi$ which
represents the gravitino. Its reduction to $d=6$ consists of four 1-form
symplectic Majorana spinors $\psi^\a= e^a\psi_a^\a$ (the gravitinos) of
both chirality and twenty symplectic Majorana spinors of both chirality
$\lambda^\a_i~(i=1,...5)$. To be more specific 
\be\ll{f1}
\hat\Psi= \o^{-{1\ov 12}} (\Psi+\sqrt{\o}\Lambda)
\ee
where
\bea\ll{f2}
&&\Psi\equiv e^a\Psi_a=e^a\psi_a - {1\ov4}e^a\g_a\tilde{\g}^i\bar{\g}\Lambda_i=
e^a\psi_a\mp {1\ov6}e^a\g_a\tilde{\g}^i\bar{\g}\lambda_i 
\br && \Lambda\equiv e^i\Lambda_i=e^i(\lambda_i - {1\ov5}\tilde{\g}_i 
\tilde{\g}^j 
\lambda_j)\pm {2\ov{15}}e^i\tilde{\g}_i\tilde{\g}^j\lambda_j
\eea
The $\o$-dependent
coefficients in \eqn{f1} are fixed by the requirement that the reduced
kinetic action will be $\o$-independent and the coefficients in \eqn{f2} are 
chosen to get the reduced kinetic action in the
standard diagonal form, that is to have

\be
\hat{\Psi}\hat\G^{(8)}D\hat\Psi=\psi\G^{(3)}\bar\g
D\psi + \lambda^i\G^a D_a\lambda_i.\ee
The spinors $\psi_a,\lambda_i$ must be split in their chiral and
antichiral components $\psi_{\pm~ a},\lambda_{\pm~i}$. All these spinors
are invariant under $SO(5,5)$. Inserting
eqs.\eqn{f1},\eqn{f2} in eq.\eqn{11} one can compute the coupling between
bosons and fermions and the four-fermion interaction terms. In particular the
interaction terms of the fermions with the scalars $\F^{uv}$, the gauge
fields $A^\otimes$ and $A_{uv}$ and the 2-form potential can be obtained from
eq.\eqn{31}. In this respect it is useful to notice that the Hodge dual of
the current $Q_7 + ^{\ast}Q_4$ of eq.\eqn{31} has the simple expression
\be
^*Q_7+Q_4={1\ov4}\hat\Psi_{\hat{b_1}}\hat\G^{[\hat{b_1}}\hat\G^{(4)}\hat 
\G^{\hat{b_2}]}\hat\Psi_{\hat{b_2}}.
\ee
As before the resulting action terms turn out to be invariant under global
transformations of $SO(5,5)$ but only local transformations of the
diagonal $SO(5)$ which corresponds to the Lorentz group of the internal
space. However, invariance under both $SO(5)$ can be recovered if one
assumes that the components of positive and negative chirality of
$\lambda_i$ are vectors of the first and the second $SO(5)$ respectively
and that $\psi^\a_{+a}$ and $\lambda^{\a}_{-\tilde{i}}$ are spinors of the first
$SO(5)$, whereas $\psi^{\dot\a}_{-a}$ and $\lambda^{\dot\a}_{+i}$ are 
spinors of the second $SO(5)$ (and of course one uses the non gauge-fixed
forms of $V_{\ud\a}^{~\ud\m}$ and  $L^{I}_{~R}$).

After a long calculation we find that the action that describes the
interaction between fermions and gauge fields and scalars can be written as
\be\ll{f3}
S_{(I)}={1\over16}\int (Q^{\a\dot\a} V_{\a\dot\a}^{~\ud\m}{\cal
F}_{\ud\m}+Q^{i\tilde{i}} P_{i\tilde{i}})
\ee
where
\bea\ll{f4}
&& ~^*Q^{\a\dot\a}=
{1\ov2}\psi_{+c}^\a\G^{[c}\G^{(2)}
\G^{d]}\psi_{-d}^{\dot{\a}}+{1\ov4} 
\psi_{+d}^\b\G^{(2)}\G^d\l_{+i}^{\dot\a} (\tilde\g^i)_\b^{~\a} 
+{1\ov4} \psi_{-d}^{\dot\b}\G^{(2)}\G^d\l_{-\tilde{i}}^{\a}
(\tilde\g^{\tilde{i}})_{\dot\b}^{~\dot\a}\br &&\hskip1.5cm
+{1\ov4} (\l_{+i}\tilde\g^{\tilde{i}})^{\dot\a}\G^{(2)}
(\l_{-\tilde{i}}\tilde\g^{{i}})^{\a} 
\br && ~^*Q^{i\tilde{i}}= \p_{+a}^\a\G^{(1)}
\G^a(\tilde\g^i\l_{-}^{\tilde{i}})_\a+
\p_{-a}^{\dot\a}\G^{(1)}\G^a(\tilde\g^{\tilde{i}}\l_{+}^{{i}})_{\dot\a}.
\eea
Action \eqn{f3} is manifestly invariant under global $SO(5,5)$ and local
$SO(5)\times SO(5)$ transformations. Moreover, we find that the splitting
in eq.\eqn{QQQ} can be performed so that 
fermionic
3-forms $Q^I$ which appear in eq.\eqn{qr} are 
\bea
&&Q^i= -{1\ov4} \psi_{+}^\a \G^{(1)}(\tilde\g^i)_\a^{~\b}\psi_{+\b} 
+{1\ov2} \psi_{-}^{\dot\a}\G^{(2)}\l_{+\dot\a}^i
-{1\ov8}\l_{-}^{~\tilde{i}\a}\G^{(3)}(\tilde\g^i)_\a^{~\b}
 \l_{-\tilde{i}\b} \br  
&&Q^{\tilde{i}}= -{1\ov4} \psi_{-}^{\dot\a}
\G^{(1)}(\tilde\g^{\tilde{i}})_{\dot\a}^{~\dot\b}\psi_{-\dot\b} 
+{1\ov2} \psi_{+}^{\a}\G^{(2)}\l_{-\a}^{\tilde{i}}
-{1\ov8}\l_{+}^{~{i}\dot\a}\G^{(3)}(\tilde\g^{\tilde{i}})_{\dot\a}^{~\dot\b}
\l_{+{i}\dot\b}   
\eea
where $Q^i = L^i_R Q^R$ and  $Q^{\tilde{i}} = L^{\tilde{i}}_R Q^R$. Then
$Q^i$, $Q^{\tilde{i}}$ are vectors of the two $SO(5)$s. It follows 
that $Q^R$, defined in eq.\eqn{qr} is an $SO(5,5)$ vector, as anticipated in
the previous section. 

In conclusion   the $d=6,~N=4$ supergravity action is\\
\bea\ll{f5}
S&=&\int\bigg[{1\ov4 }e R 
 -{1\ov 2(16)}{\cal F}_{\ud\a} ~^*{\cal
F}^{\ud\a} 
+ \eta_{RS}\left( v ~h^R~H^S+{1\ov2}dB^R (J^S +Q^S) + {1\ov{2}}J^R Q^S\right)
\br&&-{1\ov (16)} 
P_{i\tilde{i}}~^*P^{i\tilde{i}}
-{1\ov2}(\psi_{+}^\a\G^{(3)}\bar\g D\psi_{+\a} 
+\psi_{-}^{\dot\a}\G^{(3)}\bar\g D\psi_{-\dot\a})
\br&&-{1\ov2}(\l_{+}^{i\dot\a}\G^aD_a\l_{+i\dot\a} 
+\l_{-}^{\tilde{i}\a}\G^aD_a\l_{-\tilde{i}\a})\bigg]\br
&&+{1\over16} (Q^{\a\dot\a} V_{\a\dot\a}^{~\ud\m}{\cal
F}_{\ud\m}+Q^{i\tilde{i}} P_{i\tilde{i}}) + {\cal{S}}_4 
\eea
The last term ${\cal S}_4$ represents the four fermion interactions.
It can be obtained straightforwardly from the reduction of the four
fermion terms in \eqn{11}, but we have not computed it explicitly here. 
In addition to be manifestly invariant under $SO(5,5)$ duality the
action \eqn{f5} is also invariant under local supersymmetry. The
supersymmetry
transformations can be obtained easily from the reduction of eqs.(8) to
(11) and using invariance under duality. However,
 a remark is in order:
the field equations for the 2-form potentials from the action \eqn{pst}
coincide with the standard ones only after a (suitable) gauge fixing of
the symmetry $ii)$ in eq. \eqn{pst1}. Therefore, as discussed in
\cite{dal-lec-ton} - \cite{ban-ber-sor} the standard supersymmetric 
transformations of the fermions
that involve the field strength $H_u$ must be modified. The correct recipe
is to replace in them the 3-forms $H^R$ with the 3-forms
$$K^R= H^R+h^R v.$$

Concluding supersymmetry transformations are
\bea\ll{f6}
&& \d e^a= \psi_{+}^\b\G^a c_{\b\a}\ep_{+}^\a+ \psi_{-}^{\dot\b}\G^a
c_{\dot\b\dot\a}\ep_{-}^{\dot\a} \br
&& \d V_{\a\dot\a}^{~\ud\m}=
{1\ov2}V_{\b\dot\b}^{~\ud\m}(\tilde\g^i)_\a^{~\b}
(\tilde\g^{\tilde{i}})_{\dot\a}^{~\dot\b} \bigg[
(\l_{+i}\tilde\g_{\tilde{i}})^{\dot\delta} \ep_{-\dot\delta} +
(\l_{-\tilde{i}}\tilde\g_{{i}})^{\delta} \ep_{-\delta}\bigg]   \br 
&& \d {\cal A}_{(1)}^{\ud\m}=V_{\a\dot\a}^{~\ud\m}\left(
\psi_{-}^{\dot\a}
\ep_{+}^\a + \ep_{-}^{\dot\a}\psi_{+}^\a +{1\ov2} \l_{+i}^{\dot\a}\G^{(1)} 
(\ep_{+}\tilde\g^i)^\a + {1\ov2}  
(\ep_{+}\tilde\g^{\tilde{i}})^{\dot\a}\G^{(1)} \l_{-\tilde{i}}^{\a}\right) \br
&& \d B_{(2)}^R= {1\ov2}L_R^i\left( \psi_{+}^\a \G^{(1)}
(\tilde\g_i\ep_{+})_\a + \l_{+i}^{\dot\a}\G^{(2)}\ep_{-\dot\a}\right) 
+{1\ov2}\tilde{L}_R^{\tilde{i}}\left( \psi_{-}^{\dot\a} \G^{(1)}
(\tilde\g_{\tilde{i}}\ep_{-})_{\dot\a} +
\l_{-\tilde{i}}^{\a}\G^{(2)}\ep_{+\a}\right) \br
&&\hskip1.5cm +{1\ov2} \delta A^{\ud\m} (\G_R)_{{\ud\m}{\ud\n}}A^{\ud\n} 
\br
&& \d \psi_{+\a}= D\ep_{+\a}+e^a V^{~\ud\m}_{\a\dot\a}\left(
{1\ov8}\g_a^{~bc}
{\cal{F}}_{bc{}\ud\m}+{3\ov4}\g^b{\cal{F}}_{ab{}\ud\m}\right)\ep_{-}^{\dot\a}
-{1\ov4}e^aK^i_{-abc}\g^{bc}(\tilde\g_i)_\a^{~\b}\ep_{+\b} \br 
&& \d \lambda_{+i\dot\a}= {1\ov4}
P_{a\,i\tilde{i}}~\g^a(\tilde\g^{\tilde{i}}\ep_{-})_{\dot\a} +
{1\ov4}V^{\ud\m}_{\dot\a\a}{\cal{F}}_{ab\ud\m}\g^{ab}(\tilde\g_i\ep_{+})^\a
+{1\ov12} K_{-iabc}\g^{abc}\ep_{-\dot\a},
\eea
Transformations for the $\psi_{-\dot\a}$ and $\l_{-\tilde{i}\a}$ can be
obtained from the above by exchanging $+\leftrightarrow
-,~i\leftrightarrow\tilde{i},~\a\leftrightarrow\dot\a$ in
the positive chirality fields respectively.  
\vskip1cm

\begin{center}

{\large {\bf Acknowledgements}}
\end{center}
This work was supported by the European Commission TMR programme\\ 
\makebox{ERBFMRX-CT96-0045}.
H.S. was also supported in part by INFN fellowship.

\end{document}